\documentclass[11pt]{cernrep}
\usepackage{amssymb,amscd}
\usepackage[figuresright]{rotating}
\usepackage{epsfig}
\usepackage{array}
\usepackage{graphics}
\usepackage{graphicx}
\usepackage{multirow}
\usepackage{supertabular}
\usepackage{longtable}
\usepackage{dcolumn}
\usepackage{xspace}
\usepackage{tabularx}
\usepackage{color}
\usepackage{calc}
\usepackage{floatflt}
\usepackage{cite,./mcite}

\begin{document}
\title{Impact of future HERA data on the determination of proton parton distribution functions using the ZEUS NLO QCD fit}
\author{Claire Gwenlan$\dagger$, Amanda Cooper-Sarkar$\dagger$ and Christopher Targett-Adams$\ddagger$}
\institute{$\dagger${Department of Physics, Nuclear and Astrophysics Laboratory, Keble Road, Oxford. OX1 3RH. UK}\\$\ddagger${Department of Physics and Astronomy, University College London, Gower Street, London. WC1E 6BT. UK.}}
\maketitle
\begin{abstract}
The high precision and large kinematic coverage of the data from 
the HERA-I running period (1994-2000) have already allowed 
precise extractions of proton parton distribution functions. 
The HERA-II running program is now underway and is expected to 
provide a substantial increase in the luminosity collected at HERA. 
In this paper, a study is presented which investigates the 
potential impact of future data from HERA on the proton 
PDF uncertainties, within the currently planned running scenario. 
Next-to-leading order QCD predictions for inclusive jet cross sections 
at the LHC centre-of-mass energy are presented using the estimated PDFs.
Finally, the effect of a possible future measurement of the 
longitudinal structure function, $F_{\rm L}$, on the gluon 
distribution is investigated.
\end{abstract}

\section{Introduction}
Since the advent of HERA, much progress has been made in determining 
the Parton Distribution Functions (PDFs) of the proton. A 
knowledge of the PDFs is vital in order to make predictions for both 
Standard Model and beyond the Standard Model processes at hadronic 
colliders. Furthermore, they must be known as precisely as possible 
in order to maximise the discovery potential for new physics at the LHC.
HERA is now in its second stage of operation.
With the measurements that can now be expected from HERA-II, 
our knowledge of PDFs should be further improved. In this paper, 
first studies of the potential impact on the PDF uncertainties of future 
measurements from HERA, are presented. 

\section{HERA physics and kinematics}
Lepton-proton deep inelastic scattering (DIS) can proceed either 
via the neutral current (NC) interaction (through the exchange of 
a $\gamma^*$ or ${\rm Z}^0$), or via the charged current 
(CC) interaction (through the exchange of a ${\rm W}^{\pm}$). 
The kinematics of lepton-proton DIS are described 
in terms of the Bjorken scaling variable, $x$, the negative invariant 
mass squared of the exchanged vector boson, $Q^2$, and the 
fraction of energy transferred from the lepton to the hadron system, $y$. 
The three quantities are related by $Q^2=sxy$, where $s$ is 
the centre-of-mass energy squared.
\begin{table}
\begin{tabular}{llcc}
\hline
   &   &{\bf HERA-I}  &{\bf HERA-II}  \\
{\bf data sample}     & {\bf kinematic coverage}  &{\bf $\mathcal{L}$ (${\rm pb}^{-1}$)}  &{\bf $\mathcal{L}$ (${\rm pb}^{-1}$)}  \\
   &   &        &{\bf (assumed)}  \\
\hline
96-97 NC $e^+p$~\cite{epj:c21:443} &  $2.7<Q^2<30000$ ${\rm GeV}^2$; $6.3 \cdot 10^{-5}<x<0.65$    &30  &30  \\
94-97 CC $e^+p$~\cite{epj:c12:411} &  $280<Q^2<17000$ ${\rm GeV}^2$; $6.3 \cdot 10^{-5}<x<0.65$  & 48 &48\\
98-99 NC $e^-p$~\cite{epj:c28:175} &  $200<Q^2<30000$ ${\rm GeV}^2$; $0.005<x<0.65$  &  16 &350 \\
98-99 CC $e^-p$~\cite{pl:b539:197} &  $280<Q^2<17000$ ${\rm GeV}^2$; $0.015<x<0.42$  &  16 &350 \\
99-00 NC $e^+p$~\cite{hep-ex:0401003} & $200<Q^2<30000$ ${\rm GeV}^2$; $0.005<x<0.65$ &  63 &350\\
99-00 CC $e^+p$~\cite{epj:c32:16} &  $280<Q^2<17000$ ${\rm GeV}^2$; $0.008<x<0.42$    &  61 &350 \\
96-97 inc. DIS jets~\cite{pl:b547:164}  &  $125<Q^2<30000$ ${\rm GeV}^2$; $E_{\rm T}^{Breit}>8$ GeV   &  37  &500 \\
96-97 dijets in $\gamma p$~\cite{epj:c23:615} &   $Q^2 \lesssim 1$ ${\rm GeV}^2$; $E_{\rm T}^{jet1,2}>14,11$ ${\rm GeV}$   &  37  &500 \\ \hline
optimised jets~\cite{chris} &   $Q^2 \lesssim 1$ ${\rm GeV}^2$; $E_{\rm T}^{jet1,2}>20,15$ ${\rm GeV}$    &  -  &500 \\
\hline
\end{tabular}
\caption{The data-sets included in the ZEUS-JETS and HERA-II projected PDF fits. 
The first column lists the type of data and the second gives the kinematic coverage. 
The third column gives the integrated luminosities of the HERA-I measurements included in the ZEUS-JETS fit.
The fourth column gives the luminosities assumed in the HERA-II projection. 
Note that the 96-97 NC and the 94-97 CC measurements have not had their luminosity scaled for the HERA-II projection.}
\label{tab:HERAIIASSUMPTIONS}
\end{table}

At leading order (LO) in the electroweak interaction, the double 
differential cross section for the NC DIS process is given in terms 
of proton structure functions,
\begin{eqnarray}
\frac{{\rm d}^2\sigma^{\rm NC}(e^{\pm}p)}{{\rm d}x{\rm d}Q^2}= 
\frac{2\pi\alpha^2}{Q^4x} \left [ Y_+ {\rm F}_{2}(x,Q^2) - y^2{\rm F}_{\rm L}(x,Q^2) \mp Y_- 
x{\rm F}_{3}(x,Q^2) \right ]  
\label{Eqn:NC}
\vspace{-0.2cm}
\end{eqnarray}
where $Y_{\pm}=1\pm(1-y)^2$. 
The structure functions are directly related to the PDFs and 
their $Q^2$ dependence is predicted by perturbative QCD. 
In particular, ${\rm F}_2$ and $x{\rm F}_3$ depend directly on 
the quark distributions. For $Q^2 \lesssim 1000$ ${\rm GeV}^2$, ${\rm F}_2$ 
dominates the lepton-proton scattering cross section and for $x < 10^{-2}$, ${\rm F}_2$ itself is 
dominated by sea quarks while the $Q^2$ dependence is 
driven by gluon radiation. Therefore, HERA data provide vital information on the sea-quarks and gluon at low-$x$. 
At high $Q^2 \gtrsim M_{\rm Z}^2$, 
the contribution from $x{\rm F}_3$ becomes increasingly significant and gives 
information on the valence quark distributions. The longitudinal 
structure function, ${\rm F}_{\rm L}$, is directly sensitive to 
the gluon, but is only important at high-$y$. 

At LO, the CC cross sections are given by,
\begin{eqnarray}
\frac{{\rm d}^2\sigma^{\rm CC}(e^{+}p)}{{\rm d}x{\rm d}Q^2} 
= \frac{G_F^2 M_W^4}{2\pi x(Q^2+M_W^2)^2} x \left [ (\bar{u}+\bar{c})+(1-y)^2(d+s) \right ] \nonumber \\ 
\frac{{\rm d}^2\sigma^{\rm CC}(e^{-}p)}{{\rm d}x{\rm d}Q^2} 
= \frac{G_F^2 M_W^4}{2\pi x(Q^2+M_W^2)^2} x \left [ ({u}+{c})+(1-y)^2(\bar{d}+\bar{s}) \right ] 
\vspace{-0.2cm}
\end{eqnarray}
so that a measurement of the $e^+p$ and $e^-p$ cross sections 
provides information on the $d$- and $u$-valence 
quarks, respectively, thereby allowing the separation of flavour. 

The QCD scaling violations in the inclusive cross section data, 
namely the QCD Compton ($\gamma^{*}q \rightarrow gq$) and 
boson-gluon-fusion ($\gamma^{*}g \rightarrow q\bar{q}$) 
processes, may also give rise to distinct jets in the final state.
Jet cross sections therefore provide a direct 
constraint on the gluon through the boson-gluon-fusion process.
\begin{figure}[Htp]
{\includegraphics[width=15cm,height=9cm]{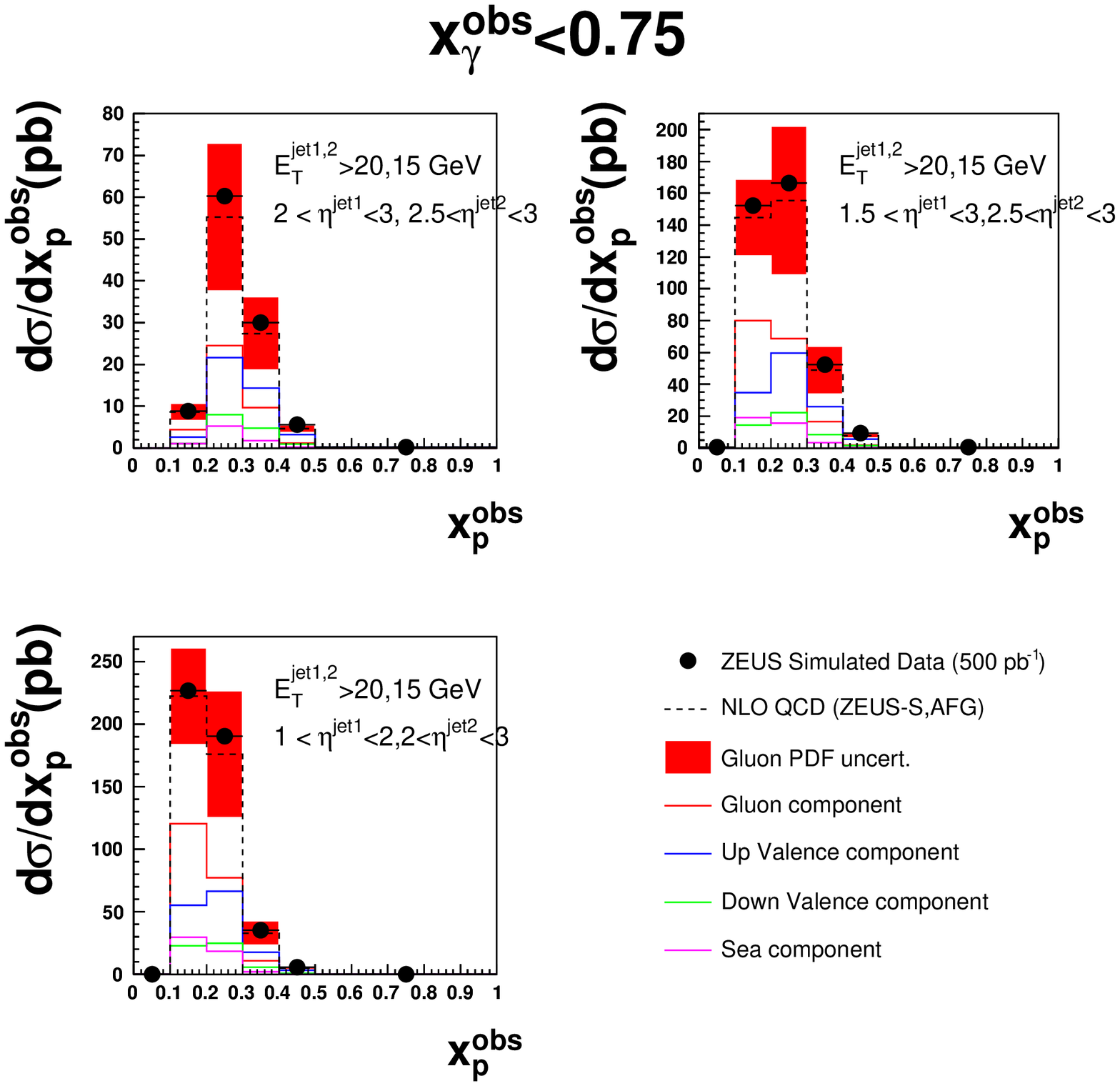}
\includegraphics[width=15cm,height=9cm]{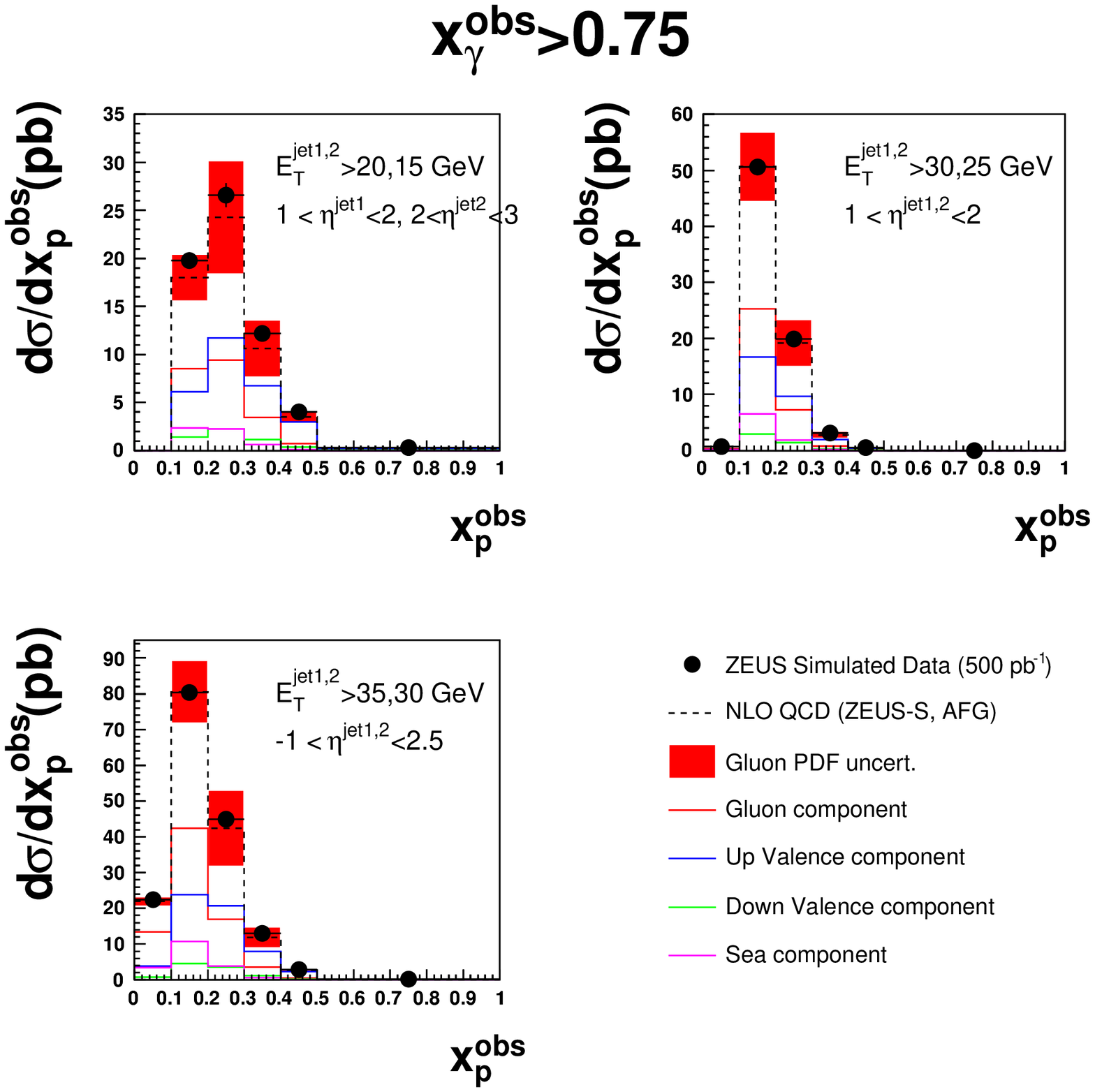}}
\caption{The optimised jet cross sections included in the HERA-II projected fit. 
The solid points show the simulated data generated using the NLO QCD programme of Frixione-Ridolfi, 
using the CTEQ5M1 proton and the AFG photon PDFs. The error bars show the statistical uncertainties, 
which correspond to $500$ ${\rm pb}^{-1}$ of HERA data. Systematic uncertainties have been neglected. 
The dashed line shows the NLO QCD prediction using the ZEUS-S proton and AFG photon PDFs. The shaded band shows the 
contribution to the cross section uncertainty arising from the uncertainty in the gluon distribution in the proton.}
\label{Fig:OptJet}
\end{figure}
\begin{figure}[Ht]
\includegraphics[width=15cm,height=9cm]{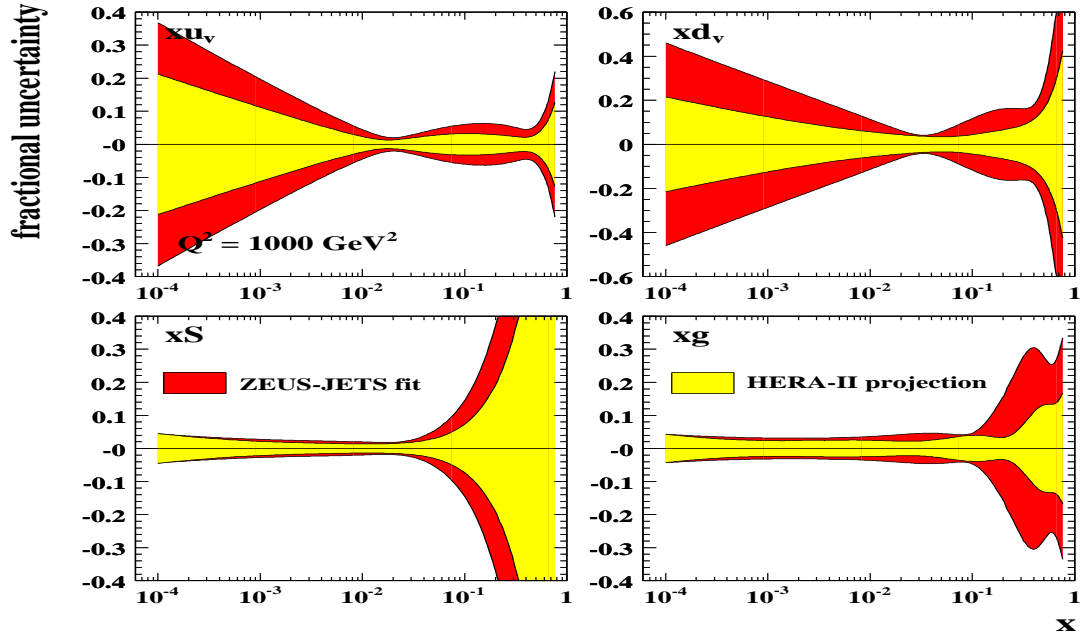}
\vspace{-0.5cm}
\caption{The fractional PDF uncertainties, as a function of $x$, for the $u$-valence, 
$d$-valence, sea-quark and gluon distributions at $Q^2=$ 1000 ${\rm GeV}^2$. 
The red shaded bands show the results of the ZEUS-JETS fit and the yellow shaded bands show the 
results of the HERA-II projected fit.}
\label{Fig:PDFS1}
\end{figure}
\begin{figure}[Ht]
\includegraphics[width=15cm,height=10.5cm]{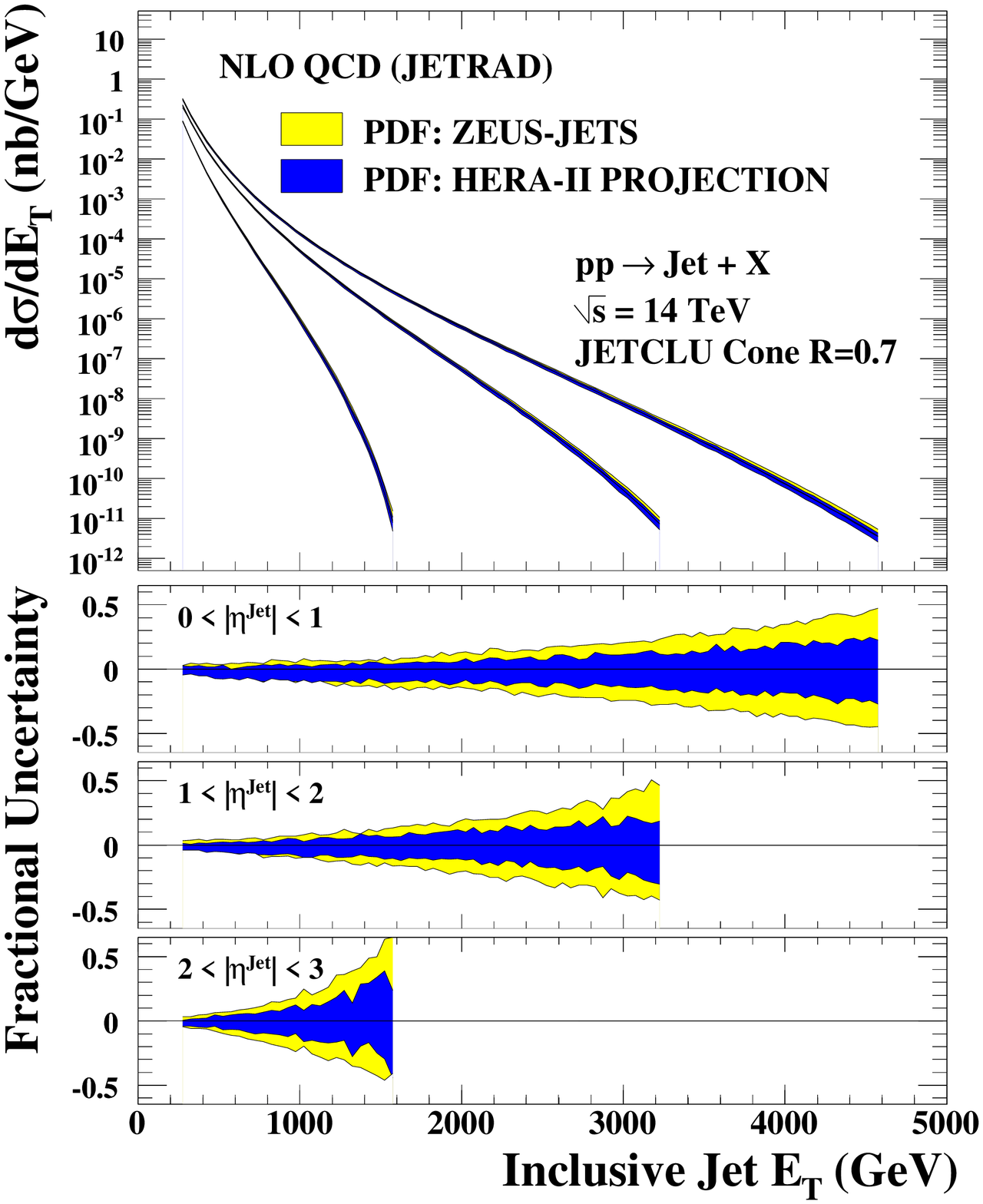}
\vspace{-0.2cm}
\caption{NLO QCD inclusive jet cross section predictions at $\sqrt{s}$=14 TeV in three 
regions of pseudo-rapidity. The yellow and blue bands show the PDF uncertainties from the ZEUS-JETS 
and HERA-II projected fits, respectively.}
\label{Fig:JETS}
\end{figure}

\vspace{-0.5cm}
\section{PDF fits to HERA data}
PDFs are usually determined in global fits~\cite{epj:c23:73,jhep:0207:012,pr:d67:012007} made within the 
conventional DGLAP~\cite{dglap} formalism at next-to-leading order (NLO). 
Such fits use data from many different experiments, with the inclusive 
cross sections from DIS experiments providing 
the most important source of information. However, the high 
precision and wide kinematic coverage of the existing HERA data 
have already allowed precise extractions of the proton PDFs using only HERA data~\cite{epj:c42:1,epj:c30:32}. 
The high statistics HERA neutral current data determine the low-$x$ 
sea and gluon distributions while information on the valence quark 
distributions is provided by the higher-$Q^2$ neutral and charged current 
data. The gluon PDF contributes only indirectly to the inclusive DIS cross section. 
However, it makes a direct contribution to jet cross sections. 

Recently, the ZEUS Collaboration have performed a 
combined NLO QCD fit to inclusive neutral and charged current 
DIS data~\cite{epj:c21:443,epj:c28:175,hep-ex:0401003,epj:c12:411,pl:b539:197,epj:c32:16} 
as well as high precision jet data in DIS~\cite{pl:b547:164} and $\gamma p$ scattering~\cite{epj:c23:615}.
This is called the ZEUS-JETS PDF fit~\cite{epj:c42:1}. 
The use of only HERA data eliminates the uncertainties from heavy-target corrections and 
removes the need for isospin symmetry assumptions. It also avoids the difficulties that can 
sometimes arise from combining data-sets from several different experiments, thereby 
allowing a rigorous statistical treatment of the PDF uncertainties. Furthermore, PDF 
uncertainties from current global fits are, in general, limited by (irreducible) 
experimental systematics. In contrast, those from fits to HERA data alone, are largely limited by 
the statistical precision of existing measurements. Therefore, the impact of future data from HERA 
is likely to be most significant in fits to only HERA data. 

\subsection{The ZEUS NLO QCD fit}
The ZEUS-JETS PDF fit has been used as the basis for all results shown in this paper. 
The most important details of the fit are summarised here. 
A full description may be found elsewhere~\cite{epj:c42:1}. 
The fit includes the full set of 
ZEUS inclusive neutral and charged current $e^{\pm}p$ data from HERA-I (1994-2000), 
as well as two sets of high precision jet 
data in $e^+p$ DIS ($Q^2 >> 1$ ${\rm GeV}^2$)
and $\gamma p$ ($Q^2 \sim 0$) scattering.
The inclusive data used in the fit, span the kinematic 
range $6.3 \times 10^{-5} < x < 0.65$ and $2.7 < Q^2 < 30 000$ ${\rm GeV}^2$. 

The PDFs are obtained by solving the NLO DGLAP equations within the $\overline{{\rm MS}}$ scheme. 
These equations yield the PDFs at all values of $Q^2$ provided they are input as functions of $x$ 
at some starting scale $Q_0^2$. The resulting PDFs are convoluted with coefficient functions to 
give predictions for structure functions and, hence, cross sections. In the ZEUS fit, the  $xu_v(x)$ ($u$-valence), 
$xd_v(x)$ ($d$-valence), $xS(x)$ (total sea-quark), $xg(x)$ (gluon) and $x(\bar{d}(x)-\bar{u}(x))$ 
PDFs are parameterised at a starting scale of $Q_0^2 = 7$ ${\rm GeV}^2$ by the form,
\begin{eqnarray}
\label{PDFparm}
xf(x) = p_{1} x^{p_2} (1-x)^{p_{3}} P(x),
\end{eqnarray}
where $P(x)=(1+p_4 x)$. No advantage in the $\chi^2$ results from using more complex polynomial forms. 
The normalisation parameters, $p_1(u_v)$ and $p_1(d_v)$, 
are constrained by quark number sum rules while $p_1(g)$ is constrained by the momentum sum rule. 
Since there is no information to constrain any difference in the low-$x$ behaviour of the $u$- and $d$-valence 
quarks, $p_{2}(u_v)$ has been set equal to $p_2(d_v)$. The data from HERA are currently less precise than the 
fixed target data in the high-$x$ regime. Therefore, the high-$x$ sea and gluon distributions 
are not well constrained in current fits to HERA data alone. 
To account for this, the sea shape has been restricted by setting $p_4(S)=0$. The high-$x$ gluon shape is constrained by 
the inclusion of HERA jet data. 
In fits to only HERA data, there is no information on the shape of $\bar{d}-\bar{u}$. Therefore, 
this distribution has its shape fixed consistent with Drell-Yan data and its normalisation 
set consistent with the size of the Gottfried sum rule violation. A suppression of the strange 
sea with respect to the non-strange sea of a factor of 2 at $Q_0^2$ is also imposed, consistent 
with neutrino induced dimuon data from CCFR. The value of the strong coupling has been fixed 
to $\alpha_s(M_{\rm Z})=0.1180$. 
After all constraints, the ZEUS-JETS fit has 11 free parameters.
Heavy quarks were treated in the variable flavour number scheme of Thorne \& Roberts~\cite{pr:d57:6871}. 
Full account was taken of correlated experimental systematic uncertainties, using the Offset Method~\cite{pr:d67:012007,jp:g28:2669}.

\section{Results and discussion}
In this section, the results of two separate studies are presented. The first study 
provides an estimate of how well the PDF uncertainties may be known by the end of HERA-II, 
within the currently planned running scenario, while the second study investigates the impact of a future HERA 
measurement of ${\rm F_L}$ on the gluon distribution. All results presented, are based on the recent 
ZEUS-JETS PDF analysis~\cite{epj:c42:1}, 
as described in the previous section.

\subsection{PDF uncertainty estimates for the end of HERA running}
The data from HERA-I are already very precise and cover a wide kinematic region.
However, HERA-II is now running efficiently and is expected to provide a substantial increase in luminosity. 
Current estimates suggest that, by the end of HERA running (in mid-2007), an integrated luminosity of $700$ ${\rm pb}^{-1}$ should be achievable. 
This will allow more precise measurements of cross sections 
that are curently statistically limited: in particular, the 
high-$Q^2$ NC and CC data, as well as high-$Q^2$ and/or high-$E_{\rm T}$ jet data. In addition to the simple increase in luminosity, 
recent studies~\cite{chris} have shown that future jet cross section measurements, in kinematic regions optimised for sensitivity to PDFs, should 
have a significant impact on the gluon uncertainties.
In this paper, the effect on the PDF uncertainties, 
of both the higher precision expected from HERA-II and the possibility of optimised jet cross section measurements,
has been estimated in a new QCD fit. This fit will be referred to as the ``HERA-II projection''. 

In the HERA-II projected fit, the statistical uncertainties on the currently available HERA-I data have been reduced.
For the high-$Q^2$ inclusive data, a total integrated luminosity of $700$ ${\rm pb}^{-1}$ was assumed, equally divided between $e^+$ and $e^-$. 
For the jet data, an integrated luminosity of $500$ ${\rm pb}^{-1}$ was assumed. The central values and systematic uncertainties were 
taken from the published data in each case. In addition to the assumed increase in precision of the measurements, 
a set of optimised jet cross sections were also included, for forward dijets in $\gamma p$ collisions, as defined in a 
recent study~\cite{chris}. Since no real 
data are yet available, simulated points were generated using the NLO QCD program of Frixione-Ridolfi~\cite{np:b507}, 
using the CTEQ5M1~\cite{cteq} proton and AFG~\cite{afg} photon PDFs. 
The statistical uncertainties were taken to correspond to $500$ ${\rm pb}^{-1}$. For this study, systematic uncertainties on 
the optimised jet cross sections were ignored. The simulated optimised jet cross section points, compared to the predictions of NLO QCD using the ZEUS-S proton PDF~\cite{zeuss}, are shown in Fig.~\ref{Fig:OptJet}. 

Table~\ref{tab:HERAIIASSUMPTIONS} lists the data-sets included in 
the ZEUS-JETS and HERA-II projected fits. The luminosities of the 
(real) HERA-I measurements and those assumed for the HERA-II projection are also given.

The results are summarised in Fig.~\ref{Fig:PDFS1}, which shows the 
fractional PDF uncertainties, for the $u$- and $d$-valence, 
sea-quark and gluon distributions, at $Q^2 = 1000$ ${\rm GeV}^2$. 
The yellow bands show the results of the ZEUS-JETS fit 
while the red bands show those for the HERA-II projection. 
Note that the same general features are observed for all values of $Q^2$. 
In fits to only HERA data, the information on the valence quarks comes from the high-$Q^2$ NC and CC cross sections. 
The increased statistical precision of the high-$Q^2$ data, as assumed in the HERA-II projected fit, 
gives a significant improvement in the valence uncertainties over the whole range of $x$. For the sea 
quarks, a significant improvement in the uncertainties at high-$x$ is also observed. In contrast, the low-$x$ 
uncertainties are not visibly reduced. This is due to the fact that the data constraining the low-$x$ region 
tends to be at lower-$Q^2$, which are already systematically limited. This is also the reason why the low-$x$ gluon 
uncertainties are not significantly reduced. However, the mid-to-high-$x$ gluon, which is constrained 
by the jet data, is much improved in the HERA-II projected fit. Note that about half 
of the observed reduction in the gluon uncertainties is due to the inclusion of the simulated optimised jet cross sections.

\vspace{-0.2cm}
\subsubsection{Inclusive jet cross sections at the LHC}
The improvement to the high-$x$ partons, observed in the HERA-II projection compared to the ZEUS-JETS fit, will be particularly 
relevant for high-scale physics at the LHC.
This is illustrated in Fig.~\ref{Fig:JETS}, which shows NLO QCD predictions from the JETRAD~\cite{np:b403:633} programme for 
inclusive jet production at $\sqrt{s}=14$ TeV. The results are shown for both 
the ZEUS-JETS and the HERA-II projected PDFs. The uncertainties on the cross sections, resulting from the PDFs, have been 
calculated using the LHAPDF interface~\cite{lhapdf}. For the ZEUS-JETS PDF, the uncertainty reaches $\sim 50\%$ at 
central pseudo-rapidities, for the highest jet transverse energies shown. 
The prediction using the HERA-II projected PDF shows a 
marked improvement at high jet tranverse energy.

\subsection{Impact of a future HERA measurement of $F_{\rm L}$ on the gluon PDF}

The longitudinal structure function, ${\rm F_L}$, is directly related to the gluon density in 
the proton.
In principle, ${\rm F}_{\rm L}$ can be extracted by measuring the NC DIS cross section at fixed $x$ and $Q^2$, 
for different values of $y$ (see Eqn.~\ref{Eqn:NC}). A precision measurement could be achieved by varying the centre-of-mass 
energy, since $s=Q^2/xy\approx4E_eE_p$, where $E_e$ and $E_p$ are the electron and proton beam energies, 
respectively. Studies~\cite{max:fl} have shown that this would be most efficiently achieved by 
changing the proton beam energy. However, such a measurement has not yet been performed at HERA.

There are several reasons why a measurement of ${\rm F_L}$ at low-$x$ could be important. 
The gluon density is not well known at low-$x$ and so 
different PDF parameterisations can give quite different predictions for ${\rm F_L}$ 
at low-$x$. Therefore, a precise measurement of the longitudinal sturcture function could both pin down the 
gluon PDF and reduce its uncertainties. Furthermore, predictions of ${\rm F_L}$ also depend upon the 
nature of the underlying theory (e.g. order in QCD, resummed calculation etc). Therefore, a measurement of 
${\rm F_L}$ could also help to discriminate between different theoretical models. 

\subsubsection{Impact on the gluon PDF uncertainties}
The impact of a possible future HERA measurement of ${\rm F_L}$ on the gluon PDF uncertainties 
has been investigated, using a set of simulated ${\rm F_L}$ data-points~\cite{max:fl}. 
The simulation was performed using the GRV94~\cite{zphys:c67:433} proton PDF for the central values, and 
assuming $E_e = 27.6$ GeV and $E_p = 920, 575, 465$ and $400$ GeV, with luminosities of 10, 5, 3 and 2 ${\rm pb}^{-1}$, respectively. 
Assuming that the luminosity scales simply as $E_p^2$, this scenario would nominally 
cost $35$ ${\rm pb}^{-1}$ of luminosity under standard HERA conditions. However, this estimate takes no account of time taken 
for optimisation of the machine with each change in $E_p$, which could be considerable. The systematic uncertainties 
on the simulated data-points were calculated assuming a $\sim 2\%$ precision on the inclusive NC 
cross section measurement. A more comprehensive description of the simulated data is given elsewhere~\cite{max:fl}.
\begin{figure}[Ht]
\includegraphics[width=15cm,height=9cm]{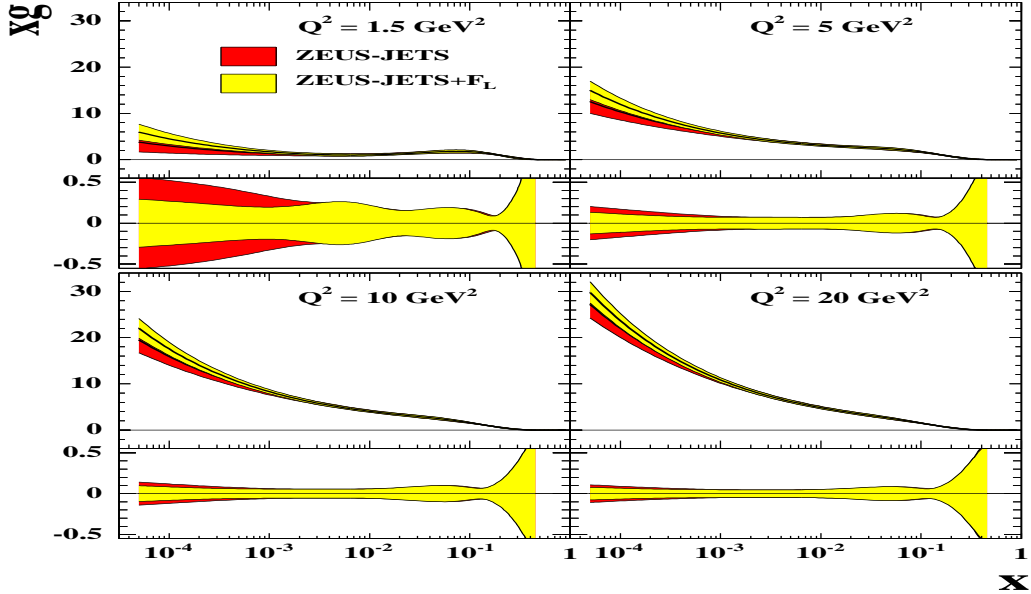}
\vspace{-0.5cm}
\caption{The gluon PDFs, showing also the fractional uncertainty, for fits with and 
without inclusion of the simulated ${\rm F_L}$ data, for $Q^2 =$ 1.5, 5, 10 and 20 
${\rm GeV}^2$. The red shaded bands show the results of the ZEUS-JETS fit and the 
yellow shaded band show the results of the ZEUS-JETS+${\rm F_L}$ fit.}
\label{Fig:FL}
\end{figure}
\begin{figure}[Ht]
\includegraphics[width=15cm,height=9cm]{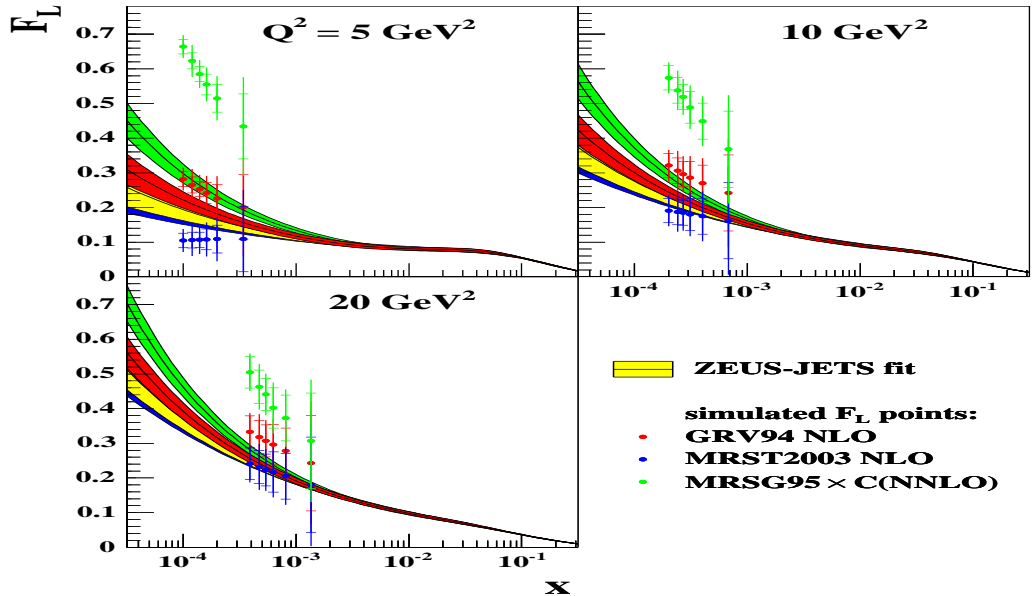}
\vspace{-0.5cm}
\caption{The distribution of the longitudinal structure function ${\rm F_L}$ at 
$Q^2=$5, 10 and 20 ${\rm GeV}^2$. 
The blue, red and green points show the simulated ${\rm F_L}$ data-points, respectively 
labelled maximum, middle and minimum in Table~\ref{tab:FLEXTREMES}. The blue, red and green shaded bands 
show the NLO QCD predictions, in the case where the data-points of the corresponding colour 
have been included in the fit. For comparison, the yellow shaded band shows 
the prediction of the ZEUS-JETS fit.}
\label{Fig:EXTREMEFL}
\end{figure}

The simulated data were included in the ZEUS-JETS fit. 
Figure~\ref{Fig:FL} shows the gluon distribution and fractional uncertainties for fits with and without 
inclusion of the simulated ${\rm F_L}$ data. The 
results indicate that the gluon uncertainties are reduced at low-$x$, but the improvement is only 
significant at relatively low $Q^2 \lesssim 20$ ${\rm GeV}^2$. 

\subsubsection{Discrimination between theoretical models}
In order to assess whether a HERA measurement of ${\rm F_L}$ could discriminate between theoretical models, 
two more sets of ${\rm F_L}$ data-points have been simulated~\cite{robert}, using different theoretical 
assumptions. The first of the two sets was generated using the MRSG95~\cite{pl:b354:155} proton PDF, which 
has a large gluon density. 
The PDFs were then convoluted with the NNLO order coefficient 
functions, which are large and positive. This gives the ``maximum'' 
set of ${\rm F_L}$ data-points. In contrast, the second set has been generated using the MRST2003~\cite{epj:c35:325} 
proton PDF, which has a negative gluon at low-$x$ and low-$Q^2$, thus providing a ``minimum'' set of 
${\rm F_L}$ data. The original set of ${\rm F_L}$ points described in the previous subsection lies 
between these two extremes. The details of all three sets are summarised in Tab.~\ref{tab:FLEXTREMES}.

Figure~\ref{Fig:EXTREMEFL} shows the results of including, individually, each set of simulated ${\rm F_L}$ 
data into the ZEUS NLO QCD fit. The results show that the NLO fit is relatively stable to the inclusion of 
the extreme sets of data. 
This indicates that a measurement of ${\rm F_L}$ could discriminate between certain theoretical models. However, it should 
be noted that the maximum and minimum models studied here were chosen specifically to give the widest possible 
variation in ${\rm F_L}$. There are many other alternatives that would lie between these extremes and the 
ability of an ${\rm F_L}$ measurement to discriminate 
between them would depend both on the experimental precision of the measurement itself, as well as the 
theoretical uncertainties on the models being tested.

\begin{table}
\begin{tabular}{lll}
\hline
  & \tablehead{1}{}{}{PDF}
  & \tablehead{1}{}{}{QCD order of coefficient functions}\\
\hline
Maximum ${\rm F_{L}}$ & MRSG95   & NNLO \\
Middle ${\rm F_{L}}$ & GRV94    & NLO\\
Minimum ${\rm F_{L}}$ & MRST2003 & NLO\\
\hline
\end{tabular}
\caption{Summary of the PDFs used to generate the simulated ${\rm F_L}$ data-points.
The extreme maximum $F_L$ points were generated using the MRSG95 PDF, and 
convoluted with NNLO coefficient functions. The middle points were generated using the GRV94 PDF, 
and the extreme minimum points were generated using the MRST2003 PDF, which has a negative gluon at low-$x$.}
\label{tab:FLEXTREMES}
\end{table}

\section{Acknowlegdements}
The authors wish to thank M. Klein and R. Thorne for providing the 
${\rm F_{L}}$ predictions, as well as for useful discussions.
C. Gwenlan wishes to thank PPARC for the support of this work.

\bibliographystyle{heralhc} 

\end{document}